\documentclass{desyproc}

\newcommand{\be}{\begin{equation}}
\newcommand{\ee}{\end{equation}}
\newcommand{\bea}{\begin{eqnarray}}
\newcommand{\eea}{\end{eqnarray}}

\newcommand{\g}{\gamma}

\newcommand{\intc}[1]{{\int\frac{d#1}{2i\pi}}}

\begin{document}
\title{Probing the BFKL dynamics at hadronic colliders}

\author{{\slshape Christophe Royon}\\[1ex]
IRFU-SPP, CEA Saclay, F91 191 Gif-sur-Yvette cedex, France
}

\contribID{smith\_joe}


\acronym{EDS'13} 

\maketitle

\begin{abstract}
We describe different possibilities to probe the BFKL dynamics at hadronic
colliders, namely Mueller-Navelet jet, and jet gap jet events. We also discuss
briefly the jet veto measurement as performed by the ATLAS collaboration at the
LHC.
\end{abstract}

\section{Forward jets and Mueller Navelet jets}

In this section, we recall briefly the previous results that we obtained
concerning forward jets at HERA and Mueller Navelet jets at the Tevatron or the
LHC as a potential test of BFKL dynamics~\cite{bfkl,fwdjet}.
Forward jets at HERA are an ideal observable to look for BFKL resummation
effects. The interval in rapidity between the scattered lepton and the jet in
the forward region is large, and when the photon virtuality $Q^2$ is close to
the transverse jet momentum $k_T$, the DGLAP cross section is small because of
the $k_T$ ordering of the emitted gluons.
The BFKL NLL formalism leads to a very good description of the forward jet
cross section measurements performed by the H1 Collaboration at 
HERA~\cite{fwdjet}, especially the triple differential cross section of forward
jet production as a function of $Q^2$, jet $p_T$ and $\xi$..

Mueller Navelet jets are another ideal processes to study BFKL resummation 
effects~\cite{mnjet}.
Two jets with a large interval in rapidity and with similar
tranverse momenta are considered. A typical observable to look for BFKL effects
is the measurement of the azimuthal correlations between both jets. The DGLAP
prediction is that this distribution should peak towards $\pi$ - ie jets
are back-to-back- whereas
multi-gluon emission via the BFKL mechanism leads to a smoother distribution.
The azimuthal correlation is an ideal variable to look for BFKL resummation
effects since it is less sensitive to experimental uncertainties such as the jet
energy scale as an example~\cite{mnjet}.
The effect of the energy conservation in the BFKL equation~\cite{mnjet} is 
large when $R$ goes
away from 1. The effect is to reduce the effective value of $\Delta \eta$ between the jets and thus the decorrelation 
effect. However, it is worth noticing that this effect is negligible when 
the ratio of the jet $p_T$s is close to 1. It is thus important to perform this
measurement as a function of the ratio of the jet $p_T$.

\section{Jet veto measurements in ATLAS}
The ATLAS collaboration measured the so-called jet veto cross
section~\cite{atlas}, namely
the events with two high $p_T$ jets, well separated in rapidity and with a veto
on jet activity with $p_T$ greater than a given threshold $Q_0$ between the two
jets. The ATLAS collaboration measured the jet veto fraction with respect to the
standard dijet cross section, and it was advocated that it might be sensitive to
BFKL dynamics. In Ref.~\cite{jetveto}, we computed the gluon emission at large
angles (which are not considered in usual MC) using the Banfi Marchesini 
Smye equation, and we showed that the
measurement can be effectively described by the gluon resummation and is thus
not related to BFKL dynamics. The sensivity to the BFKL resummation effects
appears when one looks for gaps between jets as described in the follwing
section.

\section{Jet gap jets at the Tevatron and the LHC}

In this section, we describe a new possible measurement which can probe BFKL
resummation effects and we compare our predictions with existing D0 and CDF
measurements~\cite{usb}.

\subsection{BFKL NLL formalism}

The production cross section of two jets with a gap in rapidity between them reads
\begin{equation}
\frac{d \sigma^{pp\to XJJY}}{dx_1 dx_2 dE_T^2} = {\cal S}f_{eff}(x_1,E_T^2)f_{eff}(x_2,E_T^2)
\frac{d \sigma^{gg\rightarrow gg}}{dE_T^2},
\label{jgj}\end{equation}
where $\sqrt{s}$ is the total energy of the collision,
$E_T$ the transverse momentum of the two jets, $x_1$ and $x_2$ their longitudinal
fraction of momentum with respect to the incident hadrons, $S$ the survival probability,
and $f$ the effective parton density functions~\cite{usb}. The rapidity gap
between the two jets is $\Delta\eta\!=\!\ln(x_1x_2s/p_T^2).$ 

The cross section is given by
\begin{equation}
\frac{d \sigma^{gg\rightarrow gg}}{dE_T^2}=\frac{1}{16\pi}\left|A(\Delta\eta,E_T^2)\right|^2
\end{equation}
in terms of the $gg\to gg$ scattering amplitude $A(\Delta\eta,p_T^2).$ 

In the following, we consider the high energy limit in which the rapidity gap $\Delta\eta$ is assumed to be very large.
The BFKL framework allows to compute the $gg\to gg$ amplitude in this regime, and the result is 
known up to NLL accuracy
\begin{equation}
A(\Delta\eta,E_T^2)=\frac{16N_c\pi\alpha_s^2}{C_FE_T^2}\sum_{p=-\infty}^\infty\intc{\g}
\frac{[p^2-(\g-1/2)^2]\exp\left\{\bar\alpha(E_T^2)\chi_{eff}[2p,\g,\bar\alpha(E_T^2)] \Delta \eta\right\}}
{[(\g-1/2)^2-(p-1/2)^2][(\g-1/2)^2-(p+1/2)^2]} 
\label{jgjnll}\end{equation}
with the complex integral running along the imaginary axis from $1/2\!-\!i\infty$ 
to $1/2\!+\!i\infty,$ and with only even conformal spins contributing to the sum, and 
$\bar{\alpha}=\alpha_S N_C/\pi$ the running coupling.

In this study, we performed a parametrised distribution of $d \sigma^{gg\rightarrow gg}/dE_T^2$
so that it can be easily implemented in the Herwig Monte Carlo~\cite{herwig} since performing the integral over
$\gamma$ in particular would be too much time consuming in a Monte Carlo. The implementation of the
BFKL cross section in a Monte Carlo is absolutely necessary to make a direct comparison with data.
Namely, the measurements are sensititive to the jet size (for instance, experimentally the gap size
is different from the rapidity interval between the jets which is not the case by definition in the
analytic calculation).

\begin{figure}
\epsfig{file=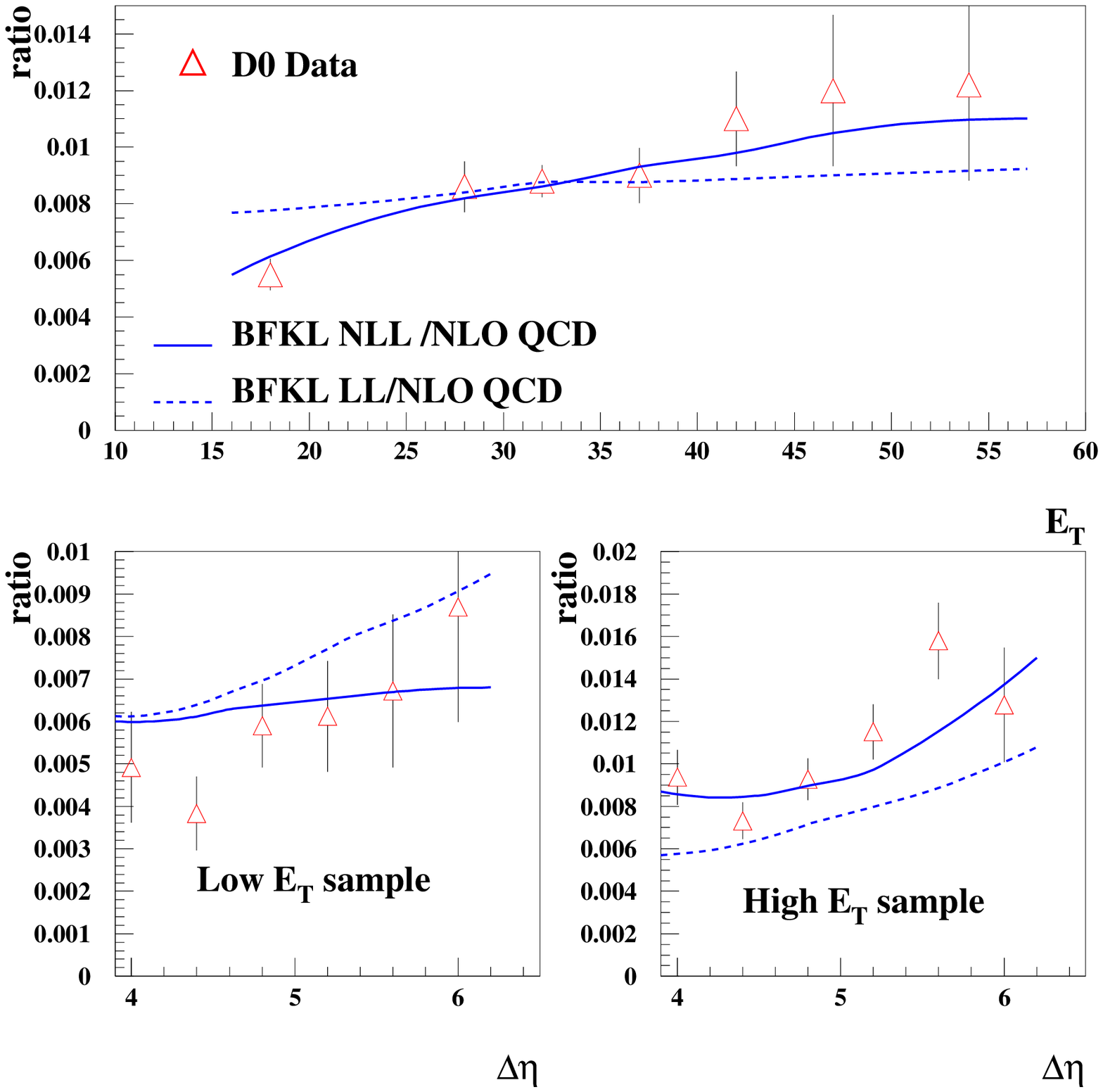,width=6.5cm}
\epsfig{file=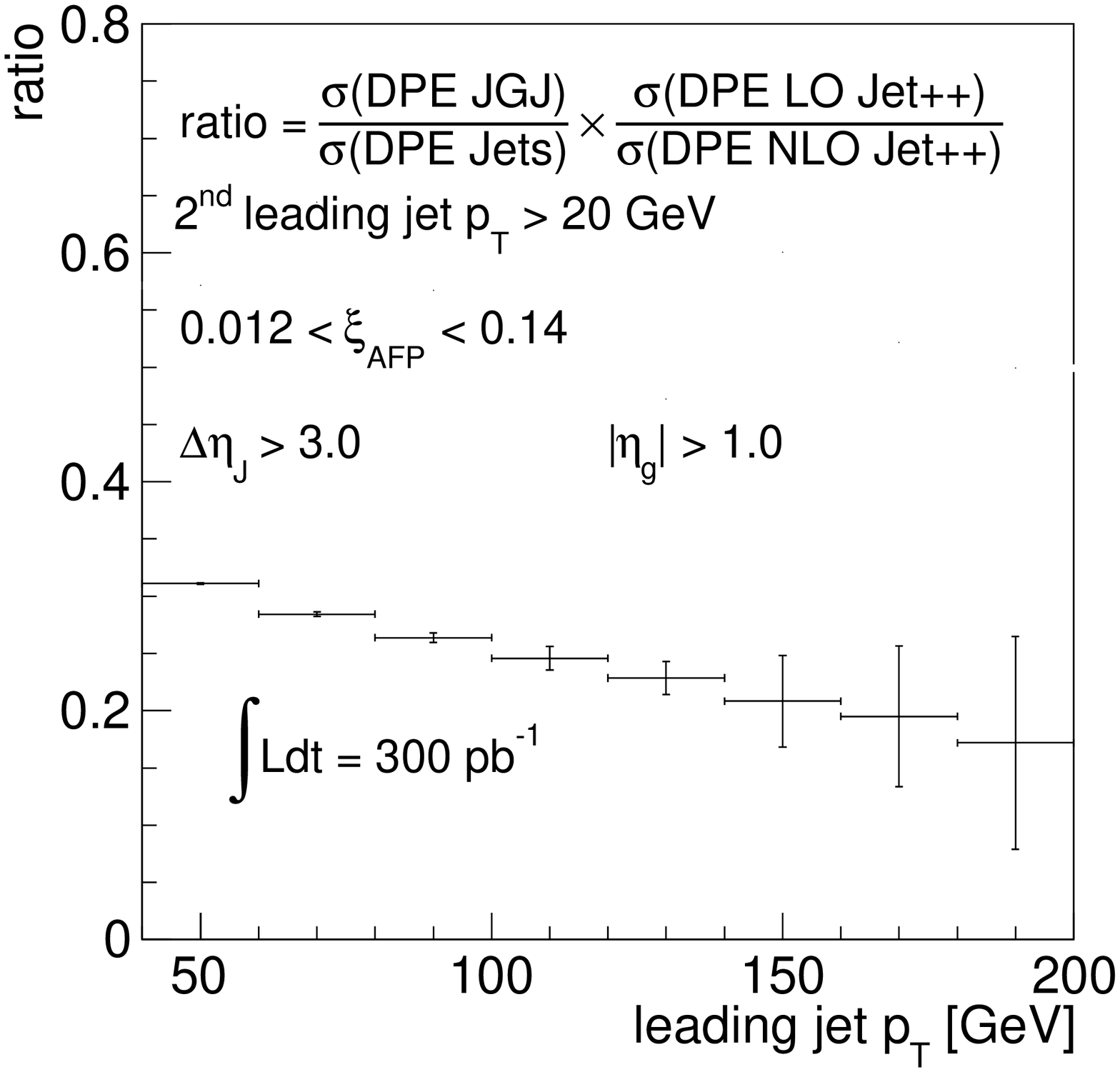,width=6.5cm}

\caption{Left: Comparisons between the D0 measurements of the jet-gap-jet event ratio with the NLL- 
and LL-BFKL calculations. The NLL calculation is in fair agreement with the data. 
The LL calculation leads to a worse description of the data.
Right: Ratio of the jet gap jet to the inclusive jet cross sections at the LHC 
as a function of jet $p_T$ in double pomeron exchange events where the protons
are detected in AFP.}
\label{d0}
\end{figure}

\subsection{Comparison with D0 and CDF measurements and predictions for LHC}
Let us first notice that the sum over all conformal spins is absolutely necessary. Considering
only $p=0$ in the sum of Equation~\ref{jgjnll} leads to a wrong normalisation and a wrong jet $E_T$
dependence, and the effect is more pronounced as $\Delta \eta$ diminishes.

The D0 collaboration measured the jet gap jet cross section ratio with respect to the total dijet
cross section, requesting for a gap between $-1$ and $1$ in rapidity, as a function of the second
leading jet $E_T$,
and $\Delta \eta$ between the two leading jets for two different low and high $E_T$ samples
(15$<E_T<$20 GeV and $E_T>$30 GeV). To compare with theory, we compute the following quantity
\begin{eqnarray}
Ratio = \frac{BFKL~ NLL~HERWIG}{Dijet~Herwig} \times \frac{LO~QCD}{NLO~QCD} 
\end{eqnarray}
in order to take into account the NLO corrections on the dijet cross
sections, where $BFKL~ NLL$ $HERWIG$ and $Dijet~Herwig$ denote the BFKL NLL and the dijet cross section
implemented in HERWIG. The NLO QCD cross section was computed using the NLOJet++ program~\cite{nlojet}.

The comparison with D0 data~\cite{usb} is shown in Fig. 1. We find a good agreement between the data
and the BFKL calculation. It is worth noticing that the BFKL NLL calculation leads to a better result
than the BFKL LL one.
The comparison with the CDF data~\cite{usb} leads to similar conclusions.

Using the same formalism, and assuming a survival probability of 0.03 at the LHC, it is possible to
predict the jet gap jet cross section at the LHC. While both LL and NLL BFKL formalisms lead to a
weak jet $E_T$ or $\Delta \eta$ dependence, the normalisation is found to be
quite different 
leading to lower cross section for the BFKL NLL formalism. The ratio of the 
jet gap jet to the inclusive jet cross sections at the LHC as a function of 
jet $p_T$ and $\Delta \eta$ is quite flat as shown in Ref.~\cite{usb}. 

\section{Jet gap jet event in diffractive processes}
A new process of detecting jet gap jet events in diffractive double pomeron
exchange processes was
introduced recently~\cite{usdiff}. The idea is to tag the intact protons inside the ATLAS
Forward Physics (AFP) detectors~\cite{loi} located at about 210 m from the ATLAS
interaction point on both sides. The advantage of such processes is that they
are quite clean since they are not ``polluted" by proton remnants and it is
possible to go to larger jet separation than for usual jet gap jet events. The
normalisation for these processes come from the fit to the D0 discussed in the
previous section. The ratio between jet gap jet to inclusive jet events is shown
in Fig. 3 requesting protons to be tagged in AFP for both samples. The ratio
shows a weak dependence as a function of jet $p_T$ (and also as a function of
the difference in rapidity between the two jets). It is worth noticing that the
ratio is about 20-30\% showing that the jet gap jet events are much more present
in the diffractive sample than in the inclusive one as expected.


\begin{footnotesize}

\end{footnotesize}

\begin{thebibliography}{99}
\bibitem{bfkl}
V.S. Fadin and L.N. Lipatov, Phys. Lett. B{\bf 429} (1998) 127;
M. Ciafaloni, Phys. Lett. B{\bf 429} (1998) 363;
M. Ciafaloni and G. Camici, Phys. Lett. B{\bf 430} (1998) 349. 

\bibitem{fwdjet} O. Kepka, C. Marquet, R. Peschanski and C. Royon, 
Phys. Lett. B{\bf 655} (2007) 236; Eur. Phys. J.
C{\bf 55} (2008) 259;   C.~Marquet and C.~Royon,
  Phys.\ Rev.\  D {\bf 79} (2009) 034028;
A. Sabio Vera and F. Schwennsen,
Nucl. Phys. B{\bf 776} (2007) 170; Phys. Rev.D
{\bf 77} (2008) 014001; A. Aktas {\it et al} [H1 Collaboration],
{\it Eur. Phys. J.} C{\bf 46} (2006) 27; H. Navelet, R. Peschanski, C. Royon,
S. Wallon, Phys.Lett. B{\bf 385} (1996) 357-364; H. Navelet, R. Peschanski,
C. Royon, Phys. Lett. B{\bf 366} (1996) 329.

\bibitem{mnjet} A.H. Mueller and H. Navelet, {\it  Nucl. Phys.} {\bf B282}
(1987) 727;
C. Marquet, C. Royon, Phys. Rev. D{\bf 79} (2009) 034028; 
B. Duclou\'e, L. Szymanowski, S. Wallon, JHEP 1305 (2013) 096.

\bibitem{atlas}  G. Aad et al, JHEP {\bf 1109}, 053 (2011).

\bibitem{jetveto} Y. Hatta, C. Marquet, C. Royon, G. Soyez, T. Ueda, D. Werder,
Phys. Rev. D{\bf87} (2013) 054016. 


\bibitem{usb} O. Kepka, C. Marquet, C. Royon, 
Phys. Rev. {\bf D83} (2011) 034036;
F.~Chevallier, O.~Kepka, C.~Marquet, C.~Royon, Phys. Rev. D{\bf 79} (2009)
094019; 
B.~Abbott {\it et al.},
  Phys.\ Lett.\  B {\bf 440} (1998) 189;
  F.~Abe {\it et al.},
  Phys.\ Rev.\ Lett.\  {\bf 80} (1998) 1156.


\bibitem{herwig} G.~Marchesini {\it et al.}, 
Comp.~Phys.~Comm.~{\bf 67} (1992) 465.
\bibitem{nlojet}
  Z.~Nagy and Z.~Trocsanyi,
  Phys.\ Rev.\ Lett.\  {\bf 87} (2001) 082001.

\bibitem{usdiff} C. Marquet, C. Royon, M. Trzebinski, R. Zlebcik, 
Phys.Rev. D{\bf 87} (2013) 3, 034010.

\bibitem{loi} ATLAS Coll., CERN-LHCC-2011-012.


\end{thebibliography}
\end{document}